\documentclass[letterpaper,english,reprint, aps,prl,superscriptaddress,showpacs]{revtex4-1}
\usepackage{textcomp}
\usepackage{amsmath}
\usepackage{amssymb}
\usepackage{graphicx}
\usepackage{subscript}

\makeatletter

\DeclareRobustCommand{\greektext}{%
  \fontencoding{LGR}\selectfont\def\encodingdefault{LGR}}
\DeclareRobustCommand{\textgreek}[1]{\leavevmode{\greektext #1}}
\DeclareFontEncoding{LGR}{}{}
\DeclareTextSymbol{\~}{LGR}{126}
\providecommand{\tabularnewline}{\\}
\newcommand{\lyxdot}{.}

\makeatother

\usepackage{babel}
\begin{document}

\title{Entanglement between Lowly and Highly Lying Atomic Spin Waves}

\author{D. S. Ding}

\email{dds@ustc.edu.cn}

\affiliation{Key Laboratory of Quantum Information, University of Science and
Technology of China, Hefei, Anhui 230026, China.}

\affiliation{Synergetic Innovation Center of Quantum Information and Quantum Physics,
University of Science and Technology of China, Hefei, Anhui 230026,
China.}

\author{K. Wang}

\affiliation{Key Laboratory of Quantum Information, University of Science and
Technology of China, Hefei, Anhui 230026, China.}

\affiliation{Synergetic Innovation Center of Quantum Information and Quantum Physics,
University of Science and Technology of China, Hefei, Anhui 230026,
China.}

\author{W. Zhang}

\affiliation{Key Laboratory of Quantum Information, University of Science and
Technology of China, Hefei, Anhui 230026, China.}

\affiliation{Synergetic Innovation Center of Quantum Information and Quantum Physics,
University of Science and Technology of China, Hefei, Anhui 230026,
China.}

\author{S. Shi}

\affiliation{Key Laboratory of Quantum Information, University of Science and
Technology of China, Hefei, Anhui 230026, China.}

\affiliation{Synergetic Innovation Center of Quantum Information and Quantum Physics,
University of Science and Technology of China, Hefei, Anhui 230026,
China.}

\author{M. X. Dong}

\affiliation{Key Laboratory of Quantum Information, University of Science and
Technology of China, Hefei, Anhui 230026, China.}

\affiliation{Synergetic Innovation Center of Quantum Information and Quantum Physics,
University of Science and Technology of China, Hefei, Anhui 230026,
China.}

\author{Y. C. Yu}

\affiliation{Key Laboratory of Quantum Information, University of Science and
Technology of China, Hefei, Anhui 230026, China.}

\affiliation{Synergetic Innovation Center of Quantum Information and Quantum Physics,
University of Science and Technology of China, Hefei, Anhui 230026,
China.}

\author{Z. Y. Zhou}

\affiliation{Key Laboratory of Quantum Information, University of Science and
Technology of China, Hefei, Anhui 230026, China.}

\affiliation{Synergetic Innovation Center of Quantum Information and Quantum Physics,
University of Science and Technology of China, Hefei, Anhui 230026,
China.}

\author{B. S. Shi}

\email{drshi@ustc.edu.cn}

\affiliation{Key Laboratory of Quantum Information, University of Science and
Technology of China, Hefei, Anhui 230026, China.}

\affiliation{Synergetic Innovation Center of Quantum Information and Quantum Physics,
University of Science and Technology of China, Hefei, Anhui 230026,
China.}

\author{G. C. Guo}

\affiliation{Key Laboratory of Quantum Information, University of Science and
Technology of China, Hefei, Anhui 230026, China.}

\affiliation{Synergetic Innovation Center of Quantum Information and Quantum Physics,
University of Science and Technology of China, Hefei, Anhui 230026,
China.}

\date{\today}
\begin{abstract}
Establishing a quantum interface between different physical systems
is of special importance for developing the practical versatile quantum
networks. Entanglement between low- and high-lying atomic spin waves
is essential for building up Rydberg-based quantum information engineering,
otherwhile be more helpful to study the dynamics behavior of entanglement
under external perturbations. Here, we report on the successful storage
of a single photon as a high-lying atomic spin wave in quantum regime.
Via storing a K-vector entanglement between single photon and lowly
lying spin wave, we thereby experimentally realize the entanglement
between low- and high-lying atomic spin waves in two separated atomic
systems. This makes our experiment the primary demonstration of Rydberg
quantum memory of entanglement, making a primary step toward the construction
of a hybrid quantum interface.
\end{abstract}


\maketitle
As a unique physical phenomenon in quantum mechanics, entanglement
entails states of two or more objects that when separated cannot be
described independently, a notion quite counterintuitive in classical
physics. It plays a vital role in quantum-information engineering
with separated entangled systems, and offers a great resource not
available within classical counterparts, and it also be facilitative
to study many fundamental quantum physics. In quantum information
science, entanglement between separated physical systems is an indispensable
resource in establishing distributed correlation across network nodes
\citep{kimble2008quantum}.

As the blockade effect of the large dipole moment of highly excited
Rydberg atom in a confined volume \citep{gaetan2009observation,urban2009observation},
a high-lying atomic spin wave from single collective Rydberg excitation
has been proposed as a potential candidate for realizing quantum computing
\citep{jaksch2000fast,lukin2001dipole}. The interacted strength between
two Rydberg atoms can be turned on and off with a contrast of 12 orders
of magnitude by preparing the atoms to Rydberg states or not \citep{saffman2010quantum},
which results in a significant advantage in realizing a C-NOT gate
\citep{isenhower2010demonstration}. Moreover, the high-lying atomic
spin wave is central to many other interesting applications such as
efficient single-photon generation \citep{dudin2012strongly}, exploration
of the attractive interaction between single photons \citep{firstenberg2013attractive},
preparation of entanglement between light and atomic excitations \citep{li2013entanglement},
all-optically switching operating using single-photon \citep{baur2014single,gorniaczyk2014single},
studying non-equilibrium phase transitions with many-body physics\citep{carr2013nonequilibrium,ding2016non}.
A low-lying atomic spin wave consisting of metastable levels is suitable
for quantum memory because of its long coherence time, a major barrier
to long-distance quantum communication \citep{duan2001long,choi2008mapping,radnaev2010quantum,kuzmich2003generation,chaneliere2005storage,zhang2011preparation,ding2015quantum,ding2015raman}.
Regarded as disparate quantum systems, connecting the low- and high-lying
atomic spin waves are crucially important in establishing long-distance
quantum communication \citep{kimble2008quantum,duan2001long} and
distributed quantum computation \citep{duan2010colloquium,van2010distributed}.
In addition, developing quantum link between low- and high-lying atomic
spin waves would make quantum networks work with superior scaling
properties and have other advantages \citep{saffman2010quantum},
such as the MHz-rate gate operations, more tolerance to some critical
parameters: weak dependence on atomic motion, independence on the
blockade shift and etc. Alternatively, such entanglement is very promising
for studying dynamics behavior of entanglement under external perturbations,
such as microwave and rf dressing. Demonstrating an entanglement between
the two is therefore interesting and merits investigation.

\begin{figure*}
\includegraphics[width=13.6cm]{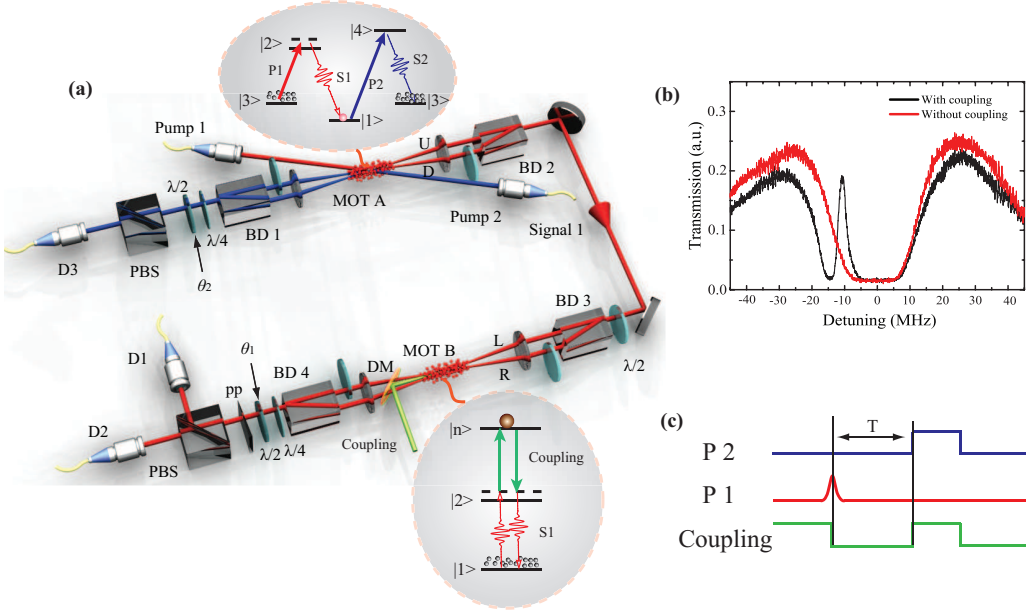}

\caption{(a) Experimental setup and energy-level diagrams. The Rubdium energy
levels dashed ellipses were used in storing signal 1 photon. \textbar{}1\textgreater{}
and \textbar{}3\textgreater{} are atomic levels of 5\textit{S}\protect\textsubscript{1/2}
(F=2) and 5\textit{S}\protect\textsubscript{1/2} (\textit{F}=3).
\textbar{}2\textgreater{} and \textbar{}4\textgreater{} are 5\textit{P}\protect\textsubscript{1/2}(\textit{F}=3)
and 5\textit{P}\protect\textsubscript{3/2}(\textit{F}=3), respectively.
\textbar{}\textit{n}\textgreater{} represents Rydberg state \textit{nD}\protect\textsubscript{3/2}.
DM: dichroic mirror. \textit{P}\protect\textsubscript{1}, \textit{P}\protect\textsubscript{2}:
pumps 1 and 2. \textit{S}\protect\textsubscript{1}, \textit{S}\protect\textsubscript{2}:
signal 1 and 2. M: mirror. BD: beam displacer. \textgreek{l}/2: half-wave
plate. \textgreek{l}/4: quarter-wave plate. \textit{pp}: the inserted
phase plate. D1,2,3: single photon detectors. \textit{\textgreek{j}}\protect\textsubscript{1,2}
is defined as the angles of the half-wave plates inserted in the paths
along which the signal 1 and signal 2 propagate, respectively. (b)
Rydberg electromagnetically induced transparency (EIT). The horizontal
axis stands for the detuning between the probe signal and the atomic
transition from 5\textit{S}\protect\textsubscript{1/2}(\textit{F}=2)
to 5\textit{P}\protect\textsubscript{1/2}(\textit{F}=3). In the experiment,
the power of the coupling laser beam is 380 mW, beam size is of $\sim$19
\textit{\textgreek{m}}m. The probe beam has a beam waist of $\sim$18
\textit{\textgreek{m}}m. (c) Time sequence for demonstrating entanglement.
T is the memory time of entanglement.}
\end{figure*}

In this letter, we report the development of a hybrid quantum link
between two distant separated atomic ensembles through exciting a
single-photon as a high-lying atomic spin wave. We first establish
the entanglement between an anti-Stokes photon and a low-lying spin
wave of one cold atomic ensemble by spontaneous Raman scattering (SRS).
Next, we send this anti-Stokes photon to excite a high-lying atomic
spin wave in another cold atomic ensemble. Via special designed interferometers,
the low- and high-lying atomic spin waves are entangled in K-vector
spaces. We demonstrate this entanglement by mapping them into two
photons and checking their entanglement. We find that the Clauser-Horne-Shimony-Holt
(CHSH) inequality is violated by more than nine standard deviations.

The medium for hybrid interface is optically thick ensembles of \textsuperscript{85}Rb
atoms trapped in two two-dimensional magneto-optical traps labeled
MOT A and MOT B (Fig. 1(a)). The temperature of the atomic cloud in
each is $\sim$200 \textit{\textgreek{m}}K and its size is 2\texttimes 2\texttimes 30
mm\textsuperscript{3} \citep{yang2012realization}. The optical depths
are 20 and 10 respectively. The hybrid quantum link involves two procedures:
a) preparing an entanglement between a single photon and the low-lying
atomic spin wave by SRS in MOT A and b) Storing single-photon as a
high-lying atomic spin wave through EIT. The experiment was run periodically
with a MOT trapping time of 7.5 ms and an experiment operating time
of 1.5 ms, which contained 3,000 operation cycles of storage, each
cycle a period of 500 ns (see time sequence in Fig. 1 (c)). Another
1 ms was used to prepare atoms to the initial atomic state $\left|3\right\rangle $
in MOT A, and state $\left|1\right\rangle $ in MOT B.

The signal-1 photon is prepared by atomic SRS process, which is correlated
with the low-lying atomic spin wave $\left|a_{low}\right\rangle {\rm {=}}{1\mathord{\left/{\vphantom{1{\sqrt{m}}}}\right.\kern -\nulldelimiterspace}{\sqrt{m}}}\sum\nolimits _{i{\rm {=1}}}^{m}e^{i\mathbf{k}_{S}\mathbf{\cdot r}_{i}}{{{\left|3\right\rangle }_{1}}\cdot\cdot\cdot{{\left|1\right\rangle }_{i}}\cdot\cdot\cdot{{\left|3\right\rangle }_{m}}}$
in $\mathbf{k}_{S}$ vector direction, where $\mathbf{k}_{S}=\mathbf{k}_{p1}-\mathbf{k}_{s1}$
is the wave vector of the low-lying atomic spin wave, $\mathbf{k}_{p1}$
and $\mathbf{k}_{s1}$ are the vectors of pump 1 and signal 1 fields,
$\mathbf{r}_{i}$ denotes the position of the $i$-th atom in atomic
ensemble. Through storing signal-1 photon through Rydberg EIT (see
Fig. 1(b)), a high-lying atomic spin wave $\left|a_{high}\right\rangle {\rm {=}}{1\mathord{\left/{\vphantom{1{\sqrt{m}}}}\right.\kern -\nulldelimiterspace}{\sqrt{m}}}\sum\nolimits _{i{\rm {=1}}}^{m}e^{i\mathbf{k}_{R}\cdot\mathbf{r}_{i}}{{{\left|1\right\rangle }_{1}}\cdot\cdot\cdot{{\left|{\rm {n}}\right\rangle }_{i}}\cdot\cdot\cdot{{\left|1\right\rangle }_{m}}}$
is realized, where $\mathbf{k}_{R}=\mathbf{k}_{C}-\mathbf{k}_{s1}$
is the wave vector of the high-lying atomic spin wave, $\mathbf{k}_{C}$
is the vector of coupling field, $\mathbf{r}_{i}$ denotes the position
of the $i$-th excited Rydberg atom in atomic ensemble. This new-type
spin wave involves a highly lying excited atom, showing a special
difference from low-lying atomic spin wave, for example the atomic
size scales as $\sim n^{2}\alpha_{0}$ ($\alpha_{0}$ is the bohr
radius, $n$ denotes the principal quantum number of Rydberg atom).
Finally, we establish the non-classical correlation between the low-
and high-lying atomic spin waves. In this process, in order to build
up the non-classical correlation between these two spin waves, small
detuning $\sim$-10 MHz (see EIT spectrum in Fig. 1(b)) is used to
match the $\sim$+10 MHz signal-1 photon. The reason to go off resonance is to reduce spontaneous emission noise in generating signal 1 field, not larger detunings is to maintain the EIT visibility. The detected signal-1 photons
before and after memory are shown in Fig. 2(a), the storage efficiency
after a programmed storage time of 300 ns is $\sim$22.9\%. In principle,
the storage efficiency can be further improved by optimizing the optical
depth of atoms, the Rabi frequency of the coupling laser, the pulse
profile of signal-1 photon and the bandwidth matching between storage
media and signal-1 photon etc.

To check whether or not the non-classical property is retained during
the storage, we map the low-lying and high-lying atomic spin waves
to the signal-1 and signal-2 photons by opening the pump 1 and coupling
pulses again after a programmed time, and check whether the Cauchy-Schwarz
inequality was violated or not \citep{kuzmich2003generation}. Classical
light satisfies $R{\rm =\left[g_{s1,s2}(t)\right]^{2}}/g_{s1,s1}(t)g_{s2,s2}(t)\le1$,
where $g_{s1,s2}(t)$ is the normalized second-order cross-correlation
between signal-1 and signal-2 photons, $g_{s1,s1}(t)$, and $g_{s2,s2}(t)$
are the corresponding auto-correlation of signal-1 and signal-2 photons
respectively. In our experiment, \textit{R}$\ge$43.2$\pm$7.3 is obtained by using the measured The auto-correlations $g_{s1,s1}(t)=1.64$, $g_{s2,s2}(t)=1.80$, the
Cauchy-Schwarz inequality was strongly violated, clearly demonstrating
the preservation of non-classical correlation during the storage of
signal 1 photon in MOT B.

The storage efficiency against storage time is shown in Fig. 2(b).
We estimate the dephasing time from Doppler decoherence is of $\sim$4.28
\textit{\textgreek{m}}s with considering the vector mismatch: \textgreek{D}\textit{k}=\textit{k}\textsubscript{475}-\textit{k}\textsubscript{795}
and the velocity of the excited Rydberg atoms of 0.276 m/s. Thus,
the Doppler decoherence and the lifetime of the Rydberg state ($n=20$,
with liftime $\sim$5\textit{\textgreek{m}}s) are not the main limitations. The additional dephasing is maybe contributed from the perturbation of external fields.

\begin{figure}
\includegraphics[width=8cm]{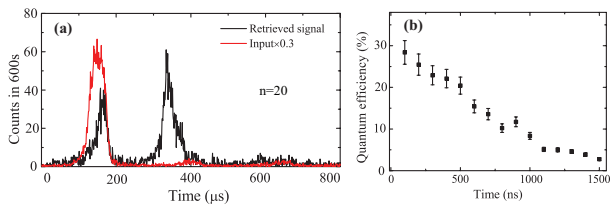}

\caption{(a) Detecting heralded signal-1 photons with storage time of 300 ns.
The storage efficiency is 22.9\%. (b) The memory efficiency vs storage
time at $n=20$.}
\end{figure}

We also characterized the single photon property of the signal 1 photon
before and after storage by checking a heralded auto-correlation parameter
${\rm g}_{{\rm s1};{\rm s1}/{\rm s2}}\left(t\right)=P_{2}P_{213}/P_{21}P_{23}$,\textit{
}which is Hanbury-Brown and Twiss (HBT) experiment on triggered signal
1 photon\citep{grangier1986experimental,chaneliere2005storage}. \textit{P}\textsubscript{2}
is the count of signal-2 photons; \textit{P}\textsubscript{21} and
\textit{P}\textsubscript{23} are the two-fold coincidence counts
between the signal-2 photons and the two separated signal-1 photons
respectively; and P\textsubscript{213} is the three-fold coincidence
counts between the signal-2 photons and the two separated signal-1
photons. A pure single photon has \textit{g}\textsubscript{s1;s1/s2}(t)=0
and a two-photon state has \textit{g}\textsubscript{s1;s1/s2}(t)=0.5.
Therefore \textit{g}\textsubscript{s1;s1/s2}(t)\textless{}1.0 violates
the classical limit and \textit{g}\textsubscript{s1;s1/s2}(t)\textless{}0.5
suggests the near-single-photon character. We obtained \textit{g}\textsubscript{s1;s1/s2}(t)
of 0.12\textpm 0.02 of the input single photons and \textit{g}\textsubscript{s1;s1/s2}(t)
is 0.10\textpm 0.01 of retrieved single photons, both closed to zero
confirmed clearly the preservation of the single-photon nature in
storage, i.e., showed definitively single high-lying atomic spin wave
in MOT B. In Refs. \citep{dudin2012strongly} and \citep{maxwell2013storage},
the input light field is a coherent light and a single high-lying
atomic spin wave is prepared via Rydberg interactions within a blockade
radius, which is confirmed by post-detecting the read-out photons.
Here, the single high-lying atomic spin wave is achieved by absorbing
heralded single photon.

\begin{table*}
\caption{Measurements of $\overline{p}_{ij}$ and concurrences $C$ before
and after collective Rydberg excitation.}

\begin{ruledtabular}
\begin{tabular}{cccccc}
 & $\overline{p}_{00}$ & $\overline{p}_{01}$ & $\overline{p}_{10}$ & $\overline{p}_{11}$ & $Con$\tabularnewline
$\overline{\rho}_{{\rm input}}$ & 0.9516$\pm$0.0008 & (2.61$\pm$0.04)\texttimes 10\textsuperscript{-2} & (2.29$\pm$0.04)\texttimes 10\textsuperscript{-2} & (2.6$\pm$0.4)\texttimes 10\textsuperscript{-5} & (3.4$\pm$0.1)\texttimes 10\textsuperscript{-2}\tabularnewline
$\rho_{{\rm output}}$ & 0.9937$\pm$0.0001 & (3.33$\pm$0.05)\texttimes 10\textsuperscript{-3} & (2.98$\pm$0.05)\texttimes 10\textsuperscript{-3} & (1.0$\pm$0.5)\texttimes 10\textsuperscript{-6} & (3.39$\pm$0.5)\texttimes 10\textsuperscript{-3}\tabularnewline
\end{tabular}\end{ruledtabular}

\end{table*}

At first, we realized the which-path entanglement of a heralded high-lying
atomic spin wave in a specially designed interferometer, which can
be written as

\begin{equation}
|{\psi_{1}}>=\frac{1}{{\sqrt{2}}}(|{0_{R}}>|{1_{L}}>+{e^{i\phi}}|{1_{R}}>|{0_{L}}>)
\end{equation}
where subscript $L$ and $R$ refer to the left and right optical
paths in the interferometer, $\phi$ denotes the relative
phase between these two optical modes, which is set to zero, and \textbar{}0\textgreater{}
and \textbar{}1\textgreater{} denote number states of high-lying atomic
spin wave, respectively. The entangled properties can be characterized
by the reduced matrix density \textit{\textgreek{r}} on the basis
of \textbar{}\textit{n}\textsubscript{\textit{L}}\textgreater{} and
\textbar{}\textit{m}\textsubscript{\textit{R}}\textgreater{} with
\{\textit{n},\textit{m}\} = \{0,1\} \citep{choi2008mapping}:
\begin{equation}
\rho{\rm {=}}\frac{1}{P}\left({\begin{array}{cccc}
{p_{00}} & 0 & 0 & 0\\
0 & {p_{10}} & d & 0\\
0 & {d^{*}} & {p_{01}} & 0\\
0 & 0 & 0 & {p_{11}}
\end{array}}\right)
\end{equation}
where $p_{ij}$ is the probability of finding \textit{i} high-lying
atomic spin waves in mode \textit{L} and \textit{j} high-lying atomic
spin waves in mode \textit{R} (see Table 1); \textit{d} is equal to
\textit{$V(p_{01}+p_{10})/2$}; and \textit{V} is the visibility of
the interference between modes \textit{L} and \textit{R} {[}see Fig.
3(b){]}. Fig. 3(a) is the input signal-1 interference between modes
\textit{L} and \textit{R}. $P$ is the total probabilities: $P=p_{00}+p_{10}+p_{01}+p_{11}$.
To characterize the entanglement properties, we use the concurrence
\citep{wootters1998entanglement} $Con{\rm {=}}\frac{1}{P}\max(0,2\left|d\right|-2\sqrt{{p_{00}}{p_{11}}})$,
which takes values between 0 and 1 representing extremes corresponding
to a separable state and a maximally entangled state. To obtain the
concurrence of the entangled state corresponding to equation (2),
we read the high-lying atomic spin wave into a single-photon state.
We measured the different probabilities, and calculated the concurrence
to be (3.39$\pm$0.5)\texttimes 10\textsuperscript{\textminus 3}
including all losses, thereby demonstrating the which-path entanglement
of a high-lying atomic spin wave. The heralded probabilities are about
3.3\texttimes 10\textsuperscript{\textminus 3} with overall optical
losses 94.6\% including photon detection loss (50\%), fiber coupling
loss (30\%), filtering losses 33.5\% (two cavity filtering loss: 30\%,
one narrowband filter loss: 5\%), two-photon excitation loss (77\%).
In principle, these losses can be reduced significantly by improving
the transmittance of the filters and the storage efficiency.

\begin{figure}
\includegraphics[width=8cm]{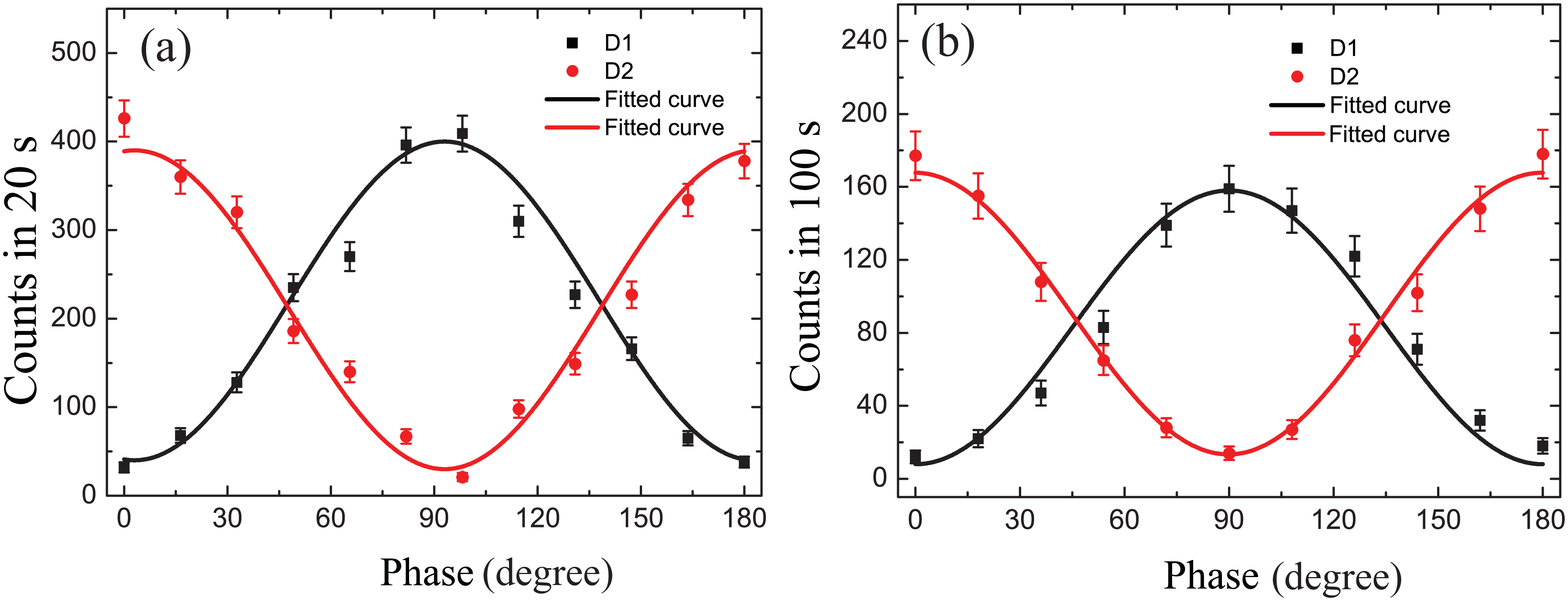}

\caption{(a) Single-photon interference between \textit{L} and \textit{R} paths.
(b) Single high-lying atomic spin wave interference with different
phases between \textit{L} and \textit{R} paths, which is controlled
by changing the phase of inserted phase plate (\textit{pp}) which
signal 1 photon passes. These counts are conditioned upon detection
of signal 2 photon in path \textit{U}. The visibilities of the interference
curves in Fig. 3(a) and (b) are 90.6\textpm 0.4\% and 85.4\textpm 0.9\%
respectively. The storage time is 300 ns.}
\end{figure}

\begin{figure}
\includegraphics[width=8.5cm]{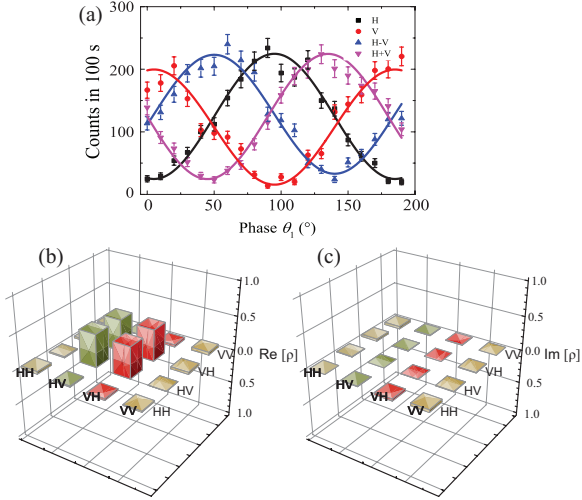}

\caption{(a)The coincidences between signal-1 and signal-2 photons against
the angle \textit{\textgreek{j}}\protect\textsubscript{1} of the
HWP\protect\textsubscript{1} through which signal 1 photon passes,
where the signal 2 photon is detected in polarization direction of
\textit{H}, \textit{V}, \textit{H}-\textit{V} and \textit{H}+\textit{V}
respectively. The interference visibilities are 84.2\%\textpm 1.0\%,
87.8\%\textpm 0.8\%, 82.3\%\textpm 1.2\%, 81.6\%\textpm 1.2\% respectively.
(b) Real and (c) imaginary parts of the density matrices of the read-out
entangled photonic state. The storage time is 300 ns.}
\end{figure}

In order to demonstrate the entanglement between low- and high-lying
atomic spin waves, we use a intrinsically stable inteferometer consisted
of two beam displacers (BD 1 and BD 2) to prepare the entanglement
between signal-1 photon and the low-lying atomic spin wave in MOT
A. In this configuration, due to the conservation of angular momentum
in SRS process, the signal-1 photons with two linearly angular momentums
(labeled as \textsl{U} and \textsl{D} directions in Fig. 1(a)) entangled
with the low-lying atomic spin waves encoded in wave vectors $\mathbf{k}_{S,U}={\mathbf{k}_{p1}}-{\mathbf{k}_{s1,U}}$
and $\mathbf{k}_{S,D}={\mathbf{k}_{p1}}-{\mathbf{k}_{s1,D}}$. The
form of the entanglement is:
\begin{equation}
|\psi_{2}>=(|U_{a}>|H_{s1}>+e^{i\varphi}|D_{a}>|V_{s1}>)/\sqrt{2}
\end{equation}
with \textit{\textgreek{f}} the relative phase between the upper and
lower optical paths, which is set as zero in our experiment, \textbar{}\textsl{U}\textsubscript{a}\textgreater{}
and \textbar{}\textsl{D}\textsubscript{a}\textgreater{} represents
the low-lying atomic spin waves encoded in wave vectors $\mathbf{k}_{S,U}$
and $\mathbf{k}_{S,D}$ respectively. \textbar{}\textsl{H}\textsubscript{s1}\textgreater{}
and \textbar{}\textsl{V}\textsubscript{s1}\textgreater{} denotes
the horizontal and vertical polarized state of signal 1 photon. We
next input the signal-1 photons into MOT B and subsequent stored it
as a high-lying atomic spin wave. With the aid of a specially designed
interferometer in MOT B, we established the entanglement between the
low-lying atomic spin wave in MOT A and the high-lying atomic spin
wave in MOT B, which can be expressed as:
\begin{equation}
|\psi_{3}>=(|U_{a}>|r_{L}>+e^{i(\varphi+\theta)}|D_{a}>|r_{R}>)/\sqrt{2}
\end{equation}
where \textbar{}\textsl{r}\textsubscript{L}\textgreater{} and \textbar{}\textsl{r}\textsubscript{R}\textgreater{}
are the corresponding state of high-lying atomic spin wave encoded
in $\mathbf{k}_{R,L}={\mathbf{k}_{C}}-{\mathbf{k}_{s1,L}}$ and $\mathbf{k}_{R,R}={\mathbf{k}_{C}}-{\mathbf{k}_{s1,R}}$
respectively.

If considering the low- and high-lying atomic spin waves individually,
the state of each spin wave are both mixed in K-vector spaces. However,
the overall state of these two spin waves cannot be described independently,
it is an entangled state. We checked this entanglement between them
by mapping the atom-atom entanglement into the photon-photon polarization
entanglement. By detecting the signal 2 photon in the polarization
direction of \textbar{}\textit{H}\textgreater{}, \textbar{}\textit{V}\textgreater{},
\textbar{}\textit{H}-\textit{V}\textgreater{}, and \textbar{}\textit{H}+\textit{V}\textgreater{}
respectively, we record the coincidence rates between signal-1 and
signal-2 photons against the angle \textit{\textgreek{j}}\textsubscript{1}
of the HWP\textsubscript{1} through which signal 1 photon passes,
and plot the two-photon interference curves (shown in Fig. 4(a)).
All visibilities are better than the threshold of 70.7\% that is the
benchmark of Bell\textquoteright s inequality, showing that entanglement
has been preserved during storage. We also used the well-known Bell-type
CHSH inequality to check the entanglement. We define the \textit{S}
value as:
\begin{equation}
S=\left|E(\theta_{1},\theta_{2})-E(\theta_{1},\theta_{2}^{\prime})+E(\theta_{1}^{\prime},\theta_{2})+E(\theta_{1}^{\prime},\theta_{2}^{\prime})\right|
\end{equation}
where \textit{\textgreek{j}}\textsubscript{1} and \textit{\textgreek{j}}\textsubscript{2}
are angles of the inserted half-wave plates shown in Fig. 1, and the
different $E({\theta_{1}},{\theta_{2}})$ are calculated using
\begin{equation}
E({\theta_{1}},{\theta_{2}})=\frac{\begin{array}{l}
C({\theta_{1}},{\theta_{2}})+C({\theta_{1}}+\frac{\pi}{2},{\theta_{2}}+\frac{\pi}{2})\\
-C({\theta_{1}}+\frac{\pi}{2},{\theta_{2}})-C({\theta_{1}},{\theta_{2}}+\frac{\pi}{2})
\end{array}}{\begin{array}{l}
C({\theta_{1}},{\theta_{2}})+C({\theta_{1}}+\frac{\pi}{2},{\theta_{2}}+\frac{\pi}{2})\\
+C({\theta_{1}}+\frac{\pi}{2},{\theta_{2}})+C({\theta_{1}},{\theta_{2}}+\frac{\pi}{2})
\end{array}}
\end{equation}
The angles of \textit{\textgreek{j}}\textsubscript{1}=0, \textit{\textgreek{j}}\textsubscript{2}=\textgreek{p}/8,
\textit{\textgreek{j}}\textsubscript{1}\textasciiacute =\textgreek{p}/4,
and \textit{\textgreek{j}}\textsubscript{2}\textasciiacute =3\textgreek{p}/8.
The \textit{S} value we obtained is 2.29\textpm 0.03. All experimental
data including two-photon visibilities and the \textit{S} value suggests
that there is an entanglement between low- and high-lying atomic spin
waves. We also performed two-qubit tomography on the read-out photons
of signal 1 and signal 2. The reconstructed density matrix (Fig. 4(a)
and Fig. 4(b)), when compared with the ideal density matrix of the
maximally entangled state, yields a calculated fidelity of 89.4\textpm 2.6\%.
We conclude again that entanglement between the low- and high-lying
atomic spin waves existed in the separated atomic ensembles.

In summary, we reported on an experiment where we have constructed
a hybrid interface between two disparate atomic systems. We have demonstrated two different entangled states in our experiment: which-path entanglement of a high-lying atomic spin wave and the entanglement between a high-lying atomic spin wave and a low-lying atomic spin wave. These two entanglement are totally different because of its¡¯ corresponding to single-particle and two-particle independently separated quantum state. The entanglement established between low- and high-lying atomic spin waves
in two atomic ensembles is physically separated 1 meter apart. With the high-lying
atomic spin wave being highly sensitive to external perturbations
such as stray electric fields and blackbody radiation, thus this hybrid
entanglement shows many prospective projects on sensing external perturbation.
Moreover, via dipole interaction between Rydberg atoms, one can in
principle demonstrate blocking or switching photonic entanglement
based on such system. Our results in establishing two atomic spin
waves with different scales show promise for advances in the field
of quantum information science and fundamental studies in quantum
physics, especially in constructing Rydberg-based quantum networks.
\begin{acknowledgments}
We very thank Wei-bin Li, Lu-Ming Duan, Zheng-Wei Zhou and Yong-Jian
Han for helpful discussions. This work was funded by the National
Natural Science Foundation of China (Grant Nos. 11174271, 61275115,
61435011, 61525504).
\end{acknowledgments}

\end{document}